\documentclass[aps, prl, amsmath, amssymb, reprint,superscriptaddress]{revtex4-1}

\usepackage{graphicx}
\usepackage{dcolumn}
\usepackage{bm}
\usepackage{subfigure}
\usepackage{enumitem}
\usepackage{hyperref}
\usepackage{color}
\usepackage{xr}
\usepackage{cleveref}
\crefname{equation}{Eq.}{Eqs.}
\crefname{figure}{Fig.}{Figs.}

\begin{document}

\title{Microphase separation in oil-water mixtures containing hydrophilic and hydrophobic ions}

\author{Nikos Tasios}
\affiliation{Soft Condensed Matter, Debye Institute for Nanomaterials Science, Utrecht University, Princetonplein 1, 3584 CC, Utrecht, The Netherlands}
\author{Sela Samin}
\author{Ren\'{e} van Roij}
\affiliation{Institute for Theoretical Physics, 
Center for Extreme Matter and Emergent Phenomena, 
Utrecht University, Princetonplein 5, 3584 CC Utrecht, The Netherlands}
\author{Marjolein Dijkstra}
\email{m.dijkstra@uu.nl}
\affiliation{Soft Condensed Matter, Debye Institute for Nanomaterials Science, Utrecht University, Princetonplein 1, 3584 CC, Utrecht, The Netherlands}

\date{\today}

\begin{abstract}    
     We develop a lattice-based Monte Carlo simulation method for charged mixtures capable of treating dielectric heterogeneities. Using this method, we study oil-water mixtures containing an antagonistic salt, with hydrophilic cations and hydrophobic anions. Our simulations reveal several phases with a spatially modulated solvent composition, in which the ions partition between water-rich and water-poor regions according to their affinity. In addition to the recently observed lamellar phase, we find tubular, droplet, and even gyroid phases reminiscent of those found in block copolymers and surfactant systems. Interestingly, these structures stem from ion-mediated interactions, which allows for tuning of the phase behavior via the concentrations, the ionic properties, and the temperature.
\end{abstract}

\maketitle

A wide variety of complex fluids displays modulated phases in equilibrium, in which spatial variations of the density or composition often originate from competing interactions \cite{seul1995}. Well-known examples with molecular-size heterogeneities include block copolymers, surfactants, and room-temperature ionic liquids. The microscopic structural features of these soft materials are crucial in applications such as catalysis, drug delivery, lithography and energy conversion \cite{schacher2012,elsabahy2012,rosen2012,weingaertner2008}. Typically, the characteristic size of spatial patterns is determined from the \emph{direct} pair interactions between the molecular constituents, e.g. the polar and apolar moieties in surfactants. Hence, the emergent structure is usually explained by the interaction mismatch between the different moieties, leading to the characteristic size of the patterns being limited to that of the molecules.

However, when the competing interactions are \emph{indirect}, modulated phases with a characteristic size much larger than that of the molecular components can be realized. In the last decade, the formation of equilibrium microheterogeneities \cite{Sadakane2007,Sadakane2011,Leys2013,Witala2015,bier2017} and ordered multilamellar structures \cite{Sadakane2009,Sadakane2013} was demonstrated in a quaternary molecular system composed of a near-critical binary solvent mixture containing a small amount of antagonistic salt, in which the cations and anions are preferentially solvated by a different solvent species. In such systems, microphase separation of the uncharged solvent components, accompanied by the partitioning of the charged species between domains, was confirmed by scattering experiments, showing a characteristic size of the microheterogeneities of the order of a few nanometers \cite{bier2017,Sadakane2009,Sadakane2013}. It has been argued that the resulting modulated phases originate from the competition between the short-range solvation of ions and long-range electrostatic forces \cite{bier2017}. 

Theoretically, such antagonistic salt solutions have been treated on the mean-field level \cite{Okamoto2010,Onuki2011,Onuki2011b,Onuki2016,bier2012,Ciach2010,Pousaneh2014}, with only a few examples in three dimensions. Although the theory has been successful in describing some of the experimental observations, it neglects fluctuations, which could be important in a near-critical system, and it considers the solvent and ionic species to be point-like, thereby neglecting excluded volume interactions. Molecular simulations of such a multi-component system are notoriously slow due to the long-range character of the Coulomb interaction and the different length-scales involved. Moreover, collective effects stemming from the dielectric inhomogeneity of the medium \cite{hirono2000second} make equilibration of the system difficult, since image charge effects have to be taken into account if one uses techniques such as the Ewald sum \cite{Arnold2005}. Efficient three-dimensional simulations are therefore needed to understand the structure and phase ordering of antagonistic salt solutions.

In this Letter, we explore a lattice model of quaternary mixtures with a new, highly efficient, Monte Carlo method which includes the complex electrostatics in polar mixtures. As in previous works \cite{Okamoto2010,Onuki2011,Onuki2011b,Onuki2016}, we find microphase separation in a wide range of solvent compositions, temperatures, and salt concentrations. However, our simulations uncover also an unexpectedly rich phase behavior in three dimensions, with several types of spatially modulated phases, one of them being the lamellar phase observed by Sadakane {\it et al} \cite{Sadakane2009,Sadakane2013}. These phases are analogous to those well known for block copolymers and surfactant systems. In our quaternary charged mixture, however, we encounter unique features in the spatial patterns, since the composition heterogeneities stem from indirect interactions, mediated by the charged species, which serve as a handle by which the structure can be controlled.

\begin{figure}
    \centering
    \includegraphics[width=0.9\columnwidth]{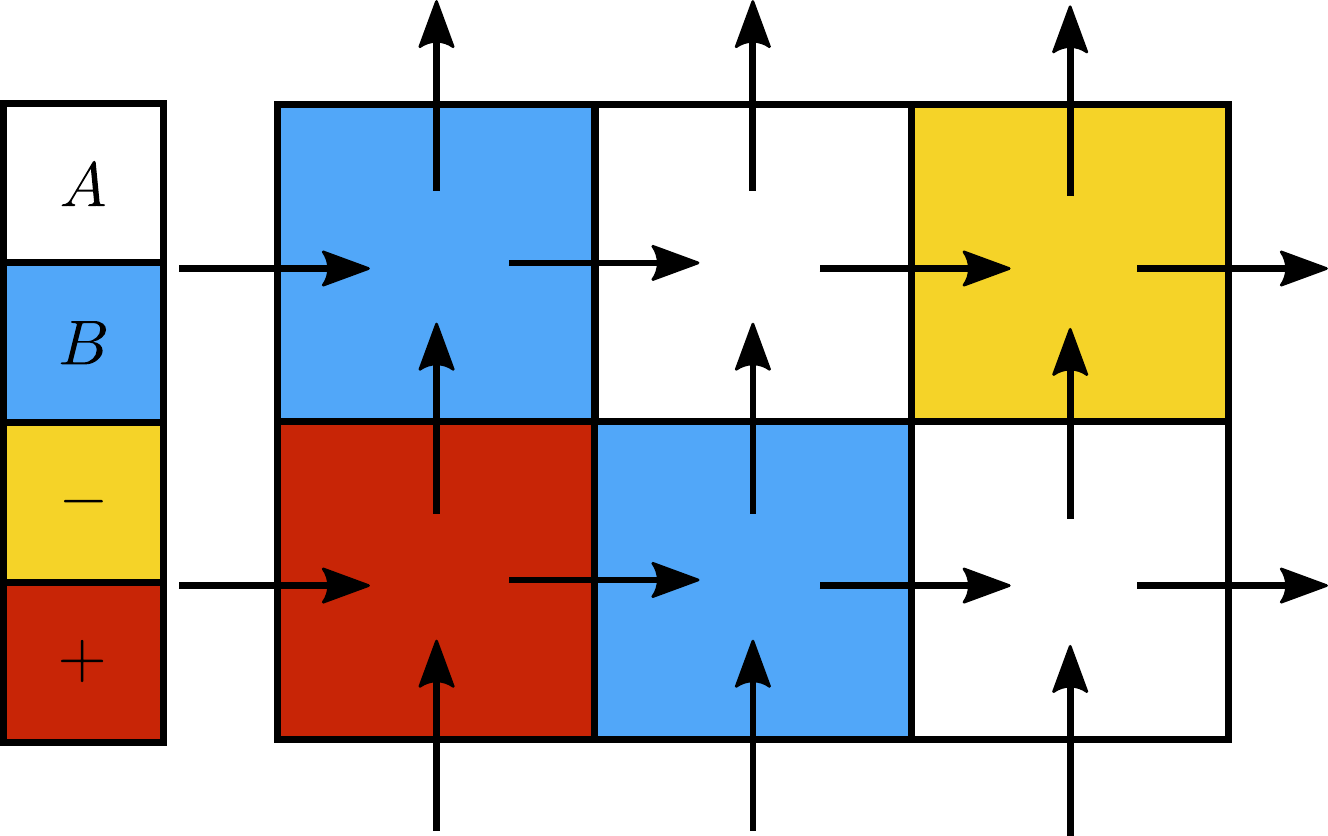}
    \caption{\label{fig:maggs_lattice}
        2D schematic representation of the lattice model for a binary AB solvent mixture with a dissolved salt consisting of $+$ and $-$ monovalent ions. The grid represents the lattice sites, and the different colors represent the different species in the system, as shown in the legend. The arrows represent the discretized electric field (not to actual scale and direction).
    }
\end{figure}

We consider a simple cubic lattice of $N=L^3$ sites of linear size $\Lambda$, our unit of length. Each lattice site $i$ can be occupied by only one of four species $\mu=A,B,+,-$, see \cref{fig:maggs_lattice} for a schematic illustration of the lattice. $A$ and $B$ are the neutral solvent species, while the ionic species, $+$ and $-$, carry a charge $\pm e$, respectively, where $e$ is the elementary charge. The incompressibility and hard-core constraint is satisfied by the occupancy operator $o_{i\mu}$, which takes a value $1$ if site $i$ is occupied by species $\mu$ and $0$ otherwise, such that $\sum_{\mu} o_{i\mu} = 1$ for $\forall i$. The dimensionless Hamiltonian of our system $\mathcal{H}/\epsilon$, where $\epsilon$ is a suitable energy scale, reads
\begin{equation}
    \label{eq:hamiltonian_full}
    \begin{split}
        \frac{\mathcal{H}}{\epsilon} = &-\frac{1}{2} \sum_{\langle i,j \rangle} \sum_{\mu , \nu} J_{\mu\nu} o_{i\mu} o_{j\nu} \\
        &+\frac{\Gamma}{4} \sum_{\langle i,j \rangle} \sum_{\mu} \frac{1}{{\varepsilon}^*_{\mu}} {D}^2_{ij}(o_{i\mu} + o_{j\mu}) ~.
    \end{split}
\end{equation}
The first term in \cref{eq:hamiltonian_full} is the short-range interaction between sites, where $\langle i,j \rangle$ denotes summation over all nearest-neighbor pairs and the interaction parameters $J_{\mu \nu}$ measure the magnitude of the interaction between species $\mu$ and $\nu$.  The second term in \cref{eq:hamiltonian_full} is the electrostatic energy, according to the method first introduced by Maggs \emph{et al.} \cite{Maggs2002,Maggs2004,Duncan2005,Levrel2005,Maggs2006,Levrel2008}, in which the (dimensionless) electric displacement field, $D_{ij}$, is discretized on the links between neighboring sites $i$ and $j$, see \cref{fig:maggs_lattice}. In \cref{eq:hamiltonian_full}, $\varepsilon^*_{\mu}$ is the reduced permittivity of species $\mu$: $\varepsilon^*_{\mu}=\varepsilon_{\mu}/\varepsilon_{A}$, with $\varepsilon_{\mu}$ and $\varepsilon_{A}$ the pure-species permittivities. This choice of permittivity units leads to the electrostatic coupling parameter $\Gamma = e^2 / (\varepsilon_A \Lambda \epsilon$) in \cref{eq:hamiltonian_full}. In this form, the parameter $\Gamma$ can be tuned (see below) to correctly capture electrostatic interactions on the lattice.

The advantage of Maggs's method is two-fold; it circumvents the time-consuming calculations of the long-ranged Coulombic interactions in charged systems, typically based on Ewald summation methods \cite{Arnold2005}, and in contrast to conventional methods, it can be straightforwardly applied to systems with a spatially varying and fluctuating dielectric permittivity (or polarization). Maggs's method uses constrained updates of an auxiliary electric displacement field instead of the electric potential, which allows the \emph{local} co-evolution of the field and the charged particles. Details of the Monte-Carlo simulation of \cref{eq:hamiltonian_full} and its derivation are given in the Supplemental Material \cite{com_sup}, with our implementation of the method also made available \cite{code}.

We choose our simulation parameters to mimic a mixture of D$_2$O (compound $B$) and 3-methylpyridine (3MP, compound $A$) with a dissolved NaBPh$_4$ salt, as in Ref. \cite{Sadakane2011}. To drive bulk phase separation in the salt-free mixture of D$_2$O-3MP, we set the nearest-neighbor interactions between the solvents to $J_{AA} = J_{AB} = 0$ and $J_{BB} = 1$. The reduced permittivity of solvent $B$ is set to $\varepsilon^*_B = 3$, which is smaller than $\varepsilon^*_B \approx 7$ in experiments. This choice increases the acceptance of Monte Carlo moves. Nevertheless, our results remain  similar for larger values of $\varepsilon^*_B$ \cite{com_sup}.

\begin{figure}
    \centering
    \includegraphics[width=0.9\columnwidth]{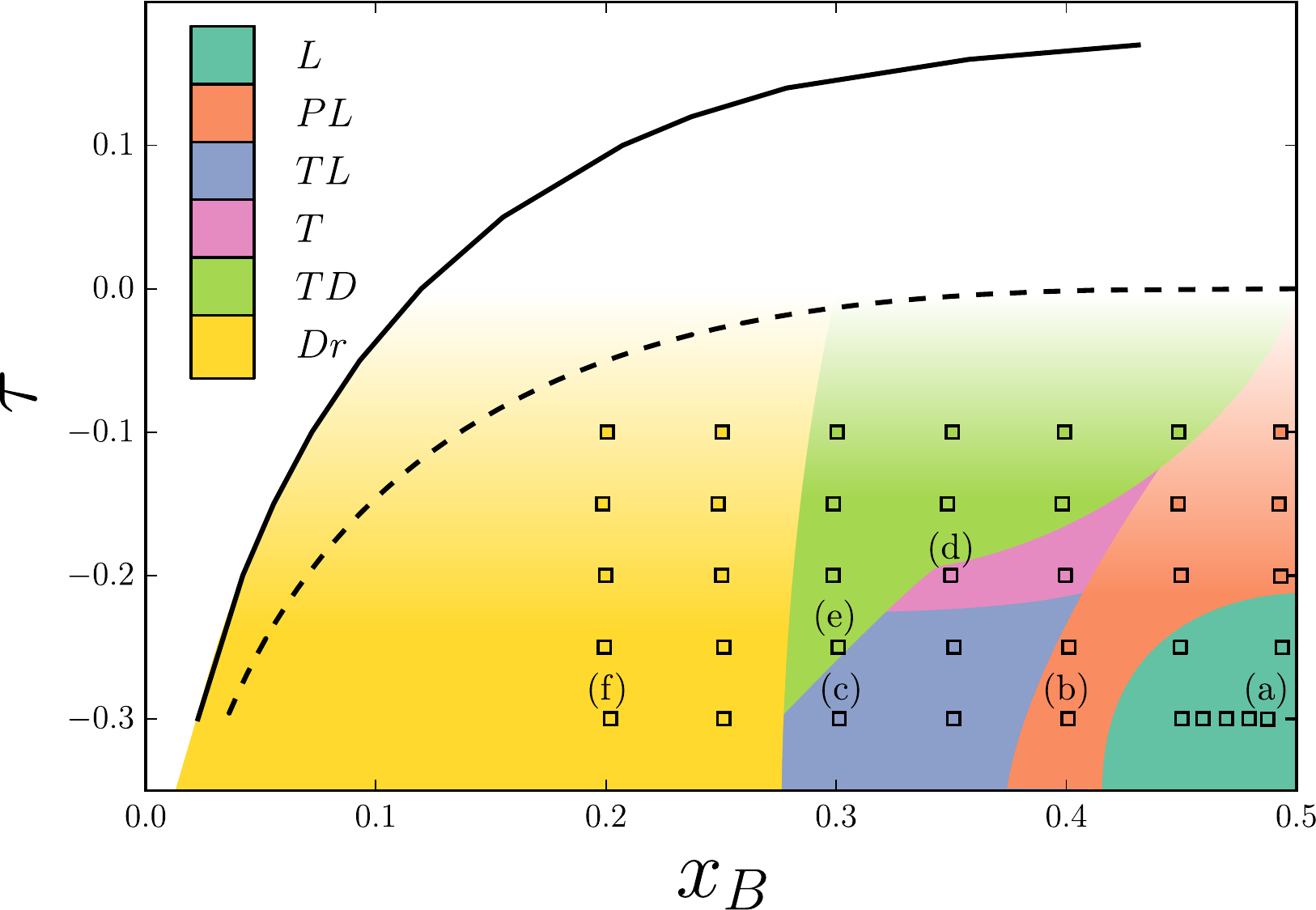}
    \caption{\label{fig:phases}
        Phase diagram of a polar $AB$ solvent mixture with dielectric contrast $\varepsilon^*_B=\varepsilon_B/\varepsilon_A=3$ (solid line), and of the lattice-gas model ($\Gamma = 0$, dashed curve), in the reduced temperature ($\tau$) and composition ($x_B$) plane. For a salt concentration of $52$ mM, the colored squares denote various mesophases found in simulations, with the color coding in the legend corresponding to the different mesophase types: lamellar (L), perforated lamellar (PL), tubular lamellar (TL), tubular (T), tubular disordered (TD), and droplet (Dr) phases. The letter markers correspond to the panels of \cref{fig:configs}. The regions of the mesophase boundaries serve only as a guide to the eye. The color gradient represents the uncertainty in the maximum temperature at which these mesophases can be detected. 
    }
\end{figure}

We first calculate the phase diagram of the polar solvent mixture, without salt, using the transition-matrix Monte Carlo (TMMC) method in the Grand-Canonical ensemble and using histogram reweighting \cite{Errington2003}. If one also ignores electrostatic effects by setting $\Gamma = 0$, \cref{eq:hamiltonian_full} reduces  to the simple lattice-gas (LG) model, that exhibits a demixing transition below a critical temperature $T^{LG}_C \approx 1.128 \epsilon / k_B$ \cite{guida1998critical}. The dashed line in \cref{fig:phases} shows the coexistence curve for this LG model in the $x_B-\tau$ plane, where $x_B$ is the fraction of $B$ lattice sites and $\tau=(T - T^{LG}_C) / T^{LG}_C$ is the reduced temperature with respect to the LG critical temperature. The solid line in \cref{fig:phases} shows the coexistence curve for the polar mixture with $\Gamma = 12$ (see below), where only the $B$-poor side, $x_B \le 0.5$, of the fairly symmetric phase diagram is shown. The demixed region of the polar mixture is broader than that of the simple LG. This is expected, since the introduction of electrostatics generates an effective (attractive) Keesom potential between same solvent sites \cite{Maggs2004}, increasing the tendency for demixing.

\begin{figure}
    \centering
    \includegraphics[width=0.95\columnwidth]{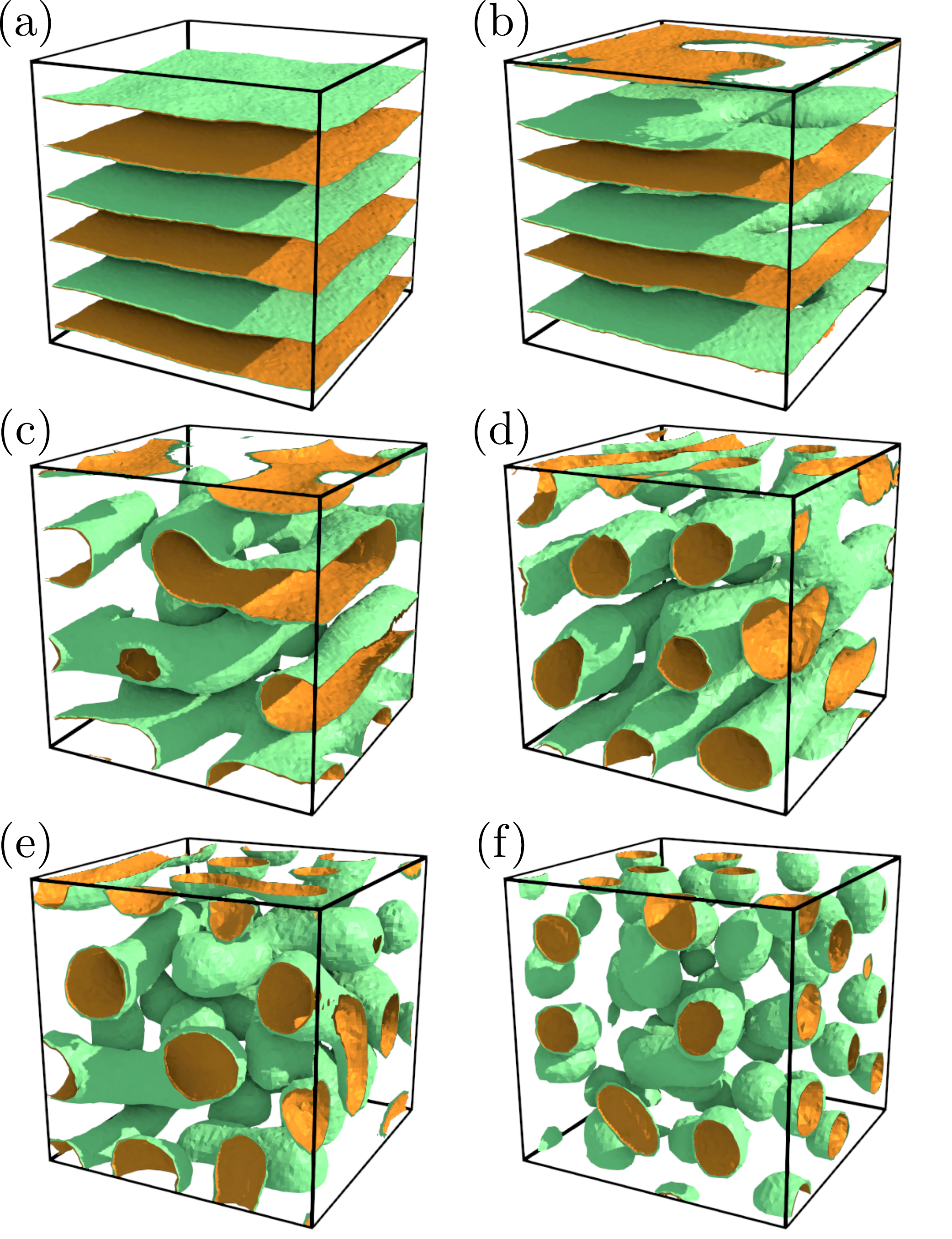}
    \caption{\label{fig:configs}
        Representation of iso-composition surfaces $x_B=0.5$. The orange/green surface represents the side of the $B$/$A$ (minority/majority) phase. We show representative state points, marked in \cref{fig:phases}, of (a) lamellar, (b) perforated lamellar, (c) tubular lamellar, (d) tubular, (e) tubular disordered, and (f) droplet phases.
    }
\end{figure}

Next, we consider the full quaternary mixture by adding to the $AB$ mixture a small amount of an antagonistic salt.
From symmetry of the Hamiltonian in \cref{eq:hamiltonian_full}, an interaction strength $J_{\pm B} = 1$ between the ions and solvent $B$ would correspond to no preferential solvation. To make the salt antagonistic, we set $J_{B+} = 6$ and $J_{B-} = -4$. Hence, the positive ions are preferentially solvated by the $B$ solvent, whereas the negative ions prefer the $A$ solvent. All other ion-solvent and ion-ion interactions are set to $0$. The above choice results in a Gibbs transfer energy (per ion at infinite dilution) from a neat B solvent to a neat A solvent, $g^\pm$, which is anti-symmetric: $g^+=-g^-\approx 15 k_B T^{LG}_C$, purely due to short-range dispersion interactions. This is a large value, but not unreasonable for highly antagonistic salts \cite{Onuki2016}.

The dielectric constants of the ions are set to $\varepsilon^*_+ = \varepsilon^*_- = 1$. Hence, an effective Keesom potential is also generated between the ions and solvent $B$, but since $\varepsilon^*_B$ is small, this only slightly increases the overall preference towards solvent $B$. The size of a lattice site is set to $\Lambda = 10${{\AA}}, equal to the hydrated size of the biggest component in the experimental system, the BPh$_4^-$ ion. Setting the lattice Bjerrum length $\lambda_B = \Gamma \epsilon \Lambda/(4\pi k_B T \varepsilon^*_A)$ to $8.4$\AA, close to the experimental value, leads to  $\Gamma = 12$ at $T^{LG}_C$. 

The quaternary mixture is simulated in the canonical ensemble, using a lattice of size $L = 64$ and a fixed number of $N_{-} = N_{+} = 4096$ ions, corresponding to an occupancy of $c_0=(N_{-}+N{+})/L^3\approxeq0.03$ or a molar concentration of $52$ mM, close to the lower limit of $60$ mM, at which an ordered-lamellar phase was experimentally observed  \cite{Sadakane2013}. All simulations with varying $x_B$ and $\tau$ are started from a random configuration of a well-mixed solution.

\begin{figure}
    \centering

\includegraphics[width=3.4in]{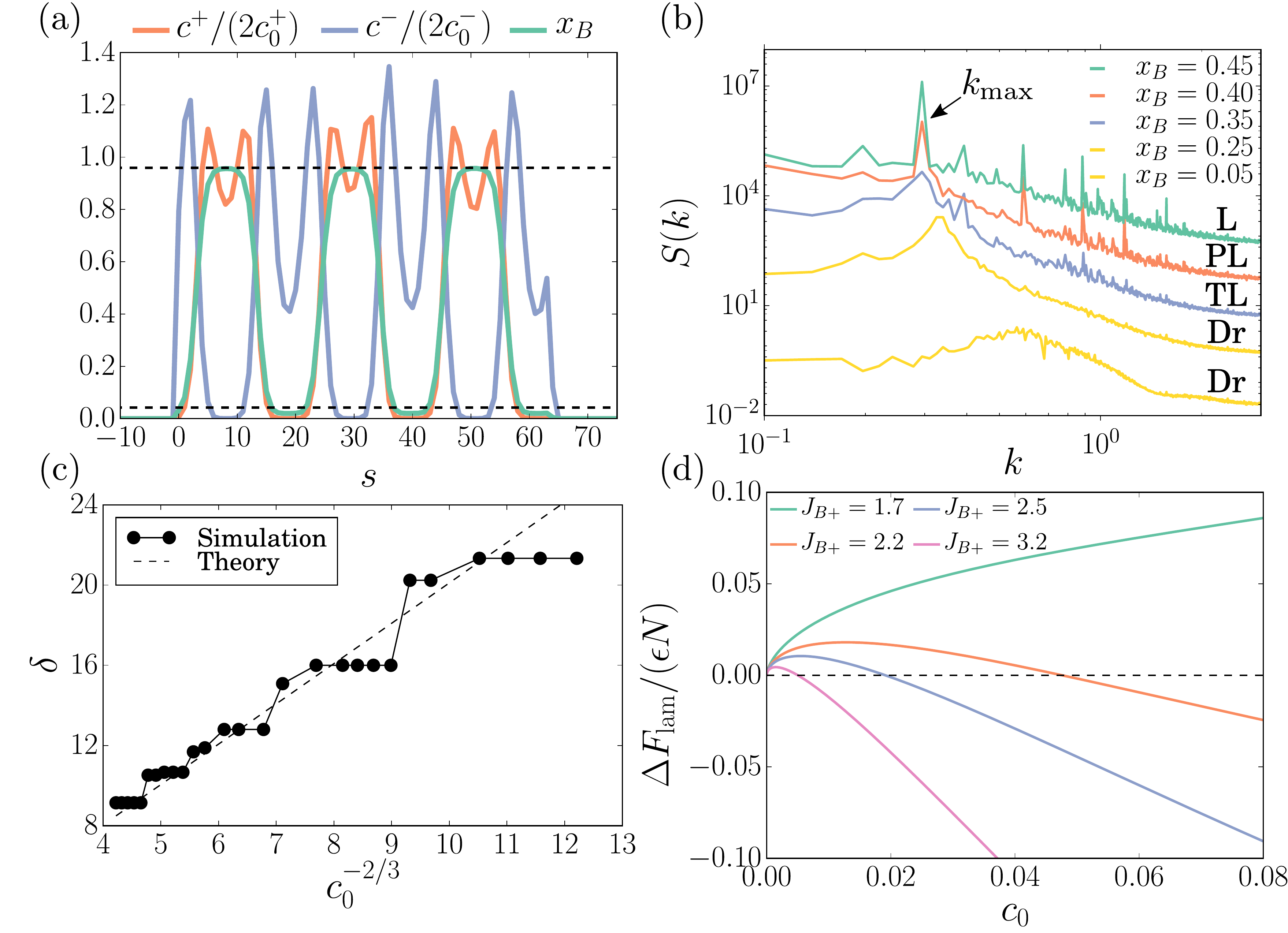}

    \caption{\label{fig:lamella_size}
        (a) $B$-solvent composition $x_B$ and the normalized ion densities, corresponding to the lamella in \cref{fig:configs}(a), as a function of the lattice position $s$ in the lamellae-normal direction. Dashed lines are the coexistence compositions of the salt-free solvent. (b) Radially averaged partial structure factor, $S_{BB}(k)$, as a function of the dimensionless wavevector, $k$, for several values of $x_B$ at $\tau = -0.3$. The corresponding phase is indicated by the boldface label.    
        (c) Lamella thickness ${\delta}$ (symbols) as a function of $c_0^{-2/3}$ for $\tau = -0.2$ and $x_B \approx 0.5$. In simulations, ${\delta}$ is calculated from ${\delta} = 2 \pi / {k}_{\text{max}}$, where ${k}_{\text{max}}$ is the position of the second largest peak in $S_{BB}(k)$, see panel (b). The dashed line corresponds to \cref{eq:ions_lamella_thickness}, where the surface tension was calculated using TMMC simulations of a salt-free mixture \cite{Errington2003}. (d) Excess lamellar free energy $\Delta F_{{\rm lam}} / (\epsilon N)$ as a function of $c_0$ at $\tau = -0.3$ and $x_B \approx 0.5$, and for several $J_{B+}$ values, with $J_{B-}=2-J_{B+}$.
    }
\end{figure}

Instead of a two-phase state, simulations of the salty mixture inside the $AB$ coexistence region reveal stable modulated mesophases with remarkably diverse structures. We visualize these structures in \cref{fig:configs}, by plotting iso-surfaces of the local composition $x_B = 0.5$. Each panel in \cref{fig:configs} corresponds to a point in the $x_B-\tau$ plane, as indicated in \cref{fig:phases}. In \cref{fig:configs}(a), we show a lamellar phase (L) of alternating composition regions, similar to the experimental observations \cite{Sadakane2009,Sadakane2013}. The same configuration is obtained from simulations that are started from a two-phase state with the salt-free coexistence compositions. \cref{fig:configs}(b) shows a perforated lamellar (PL) phase, and \cref{fig:configs}(c) a tubular lamellar (TL) phase, with lamella showing a tube-like structure, where one could argue that the TL phase is actually an extreme case of the PL phase. In \cref{fig:configs}(d)-(e), we also identify an ordered tubular (T) phase, where the minority solvent is organized in parallel tubes, and a tubular disordered (TD) phase, where the tubes are disordered. Lastly, a disordered droplet (Dr) phase of the minority component is found for low enough compositions $x_B$, see \cref{fig:configs}(f). A partial structure factor $S_{BB}(k)$ \cite{hansen} of all the different microphases obtained from simulations \cite{com_sup,Newman1999} is presented in \cref{fig:lamella_size}(b) and in \cite{com_sup}, revealing (i) multiple peaks not only for L but also for PL and even TL phases, and (ii) a single broad peak for the T, TD and Dr phases. Therefore, it is possible that some of these phases could have gone unnoticed in scattering experiments \cite{Sadakane2007,Sadakane2011,Leys2013,Witala2015,bier2017}.

A summary of all the state points we investigated and their classification is shown in \cref{fig:phases}, which shows that droplets form for $x_B \lesssim 0.25$ and lamellar and tubular structures for $x_B \gtrsim 0.25$. For temperatures $\tau \lesssim -0.2$, we find lamellar-like phases, of which only the perforated lamellar phase exists at higher temperatures. The system transitions from the lamellar phase to the tubular-lamellar phase by reducing the fraction of $B$ solvent, which dictates more compact $B$-rich domains. For higher temperatures, $\tau \gtrsim 0.1$, the system becomes disordered, exhibiting a bicontinuous structure \cite{com_sup} at first, and eventually becoming fully mixed at high enough $\tau$. We stress that all the mesophases were also stable at smaller values of $J_{\pm B}$, and in some cases with $\varepsilon^*_B =1$. Moreover, for $\varepsilon^*_B = 9$, which is close to the experimental value, we found  additional phases such as a gyroid phase and hexagonally-ordered droplet and tubular phases \cite{com_sup}. By contrast, in three-dimensional simulations of a mean-field model \cite{Onuki2016}, only bicontinuous and tube-like domains have been found until now. 

The detailed structure of the lamellar phase is presented in \cref{fig:lamella_size}(a), where we plot profiles of $B$ solvent and ion compositions, corresponding to the state of \cref{fig:configs}(a), as a function of the lattice position $s$ in the lamella-normal direction. The figure shows that the composition in the middle of the lamella (dashed lines) approaches a salt-free solvent mixture as the ions almost completely partition between the lamellae, with almost all the positive (negative) ions in the $B$-rich ($A$-rich) regions. The ion concentration is, however, higher at the lamella interfaces, where a back-to-back electric double layer is formed. The slight asymmetry between the ion profile in each phase stems from the effective Keesom potential that increases the affinity for solvation of both ionic species in the $B$ solvent.

We propose a simplified mean-field model for the lamellar phase formation, since this phase is relatively easy to analyze and can be related to experimental findings. The structure revealed by \cref{fig:lamella_size}(a) suggests that, as a first approximation, we may assume that (i) both ionic species and solvents partition completely between the lamellae, (ii) the species densities depend weakly on position within the lamellae, and (iii) the (dimensionless) surface tension, $\gamma$, at the lamellar interfaces is not too much affected by the presence of ions \cite{Onuki2016}. We therefore treat the system as oppositely charged slabs with alternating composition. The resulting free-energy difference between the lamellar and demixed two-phase states $\Delta F_{{\rm lam}}$ is given in the Supplemental Material \cite{com_sup}. Minimization of $\Delta F_{{\rm lam}}$ w.r.t. the lamella thickness $\delta$ yields
\begin{equation}
    {\delta} = \left(\frac{192{\gamma} }{\Gamma c_0^2 \left[ x_B/\varepsilon^*_B + (1 - x_B) \right]} \right)^{1/3}\propto c_0^{-2/3}~.
\label{eq:ions_lamella_thickness}
\end{equation}
We test \cref{eq:ions_lamella_thickness} by plotting the simulated thickness of the lamellae against $c_0^{-2/3}$ in \cref{fig:lamella_size}(c).
There is a good quantitative agreement between \cref{eq:ions_lamella_thickness} and the simulation results, although in the simulation ${\delta}$ changes step-wise, with increasingly larger steps, due to the finite size of the simulation box. Similar to the results of Ref. \cite{Sadakane2013}, the lamellae thickness $\delta$ increases with decreasing salt concentration. Our model highlights the important interplay between electrostatics and surface tension in forming lamellae. However, the model is far too simplistic for real systems where the ion partitioning is partial and, more importantly, molecular size asymmetry plays a significant role in structure formation \cite{Sadakane2013,Witala2015}.

Putting $\delta$ from \cref{eq:ions_lamella_thickness} back into the free energy $\Delta F_{{\rm lam}}$, we can estimate when the lamellar phase is favored ($\Delta F_{{\rm lam}}<0$) over demixed two-phase states. In \cref{fig:lamella_size}(d) we plot $\Delta F_{{\rm lam}}$ for several $J_{B+}$ values as a function of the salt concentration. The lamellar phase is favored only above a critical salt concentration for large enough $J_{B+}$, such that $\Delta F_{{\rm lam}}=0$ exists, which is also confirmed by simulations. The critical concentration decreases with increasing $J_{B+}$, since this favors ion partitioning and hence lamellae formation.

In conclusion, we performed three-dimensional Monte-Carlo simulations of binary oil-water mixtures containing antagonistic salts, which point towards the possible existence of more mesophases than observed thus far. Since only the lamellar phase has been characterized until now, we hope that our findings will motivate further experimental work to explore the D$_2$O -3MP-NaBPh$_4$ system and others in more depth. In near-critical conditions, however, mesophase fluctuations become large and therefore larger simulation boxes are needed to determine critical behavior. Work to significantly increase the simulated domain by parallelizing our code is underway. We hope that this will enable us to shed light on the critical features of the mesophases in the future. Although our lattice model was constructed specifically for a charged quaternary mixture, it could straightforwardly be extended to study mesoscale systems of other charged complex fluids, for example, the challenging problem of polymeric complex coacervation \cite{sing2017}.

\begin{acknowledgments}
N.T. and M.D. acknowledge financial support from an NWO-ECHO grant.
R.v.R acknowledges financial support of a Netherlands 
Organisation for 
Scientific Research (NWO) VICI grant funded by the Dutch Ministry
of Education, Culture and Science (OCW). S.S acknowledges funding from the 
European Union's Horizon 2020 research and innovation programme under the Marie 
Sk\l{}odowska-Curie grant agreement No. 656327. This work is part of the D-ITP 
consortium, a program of NWO funded by OCW.
\end{acknowledgments}

\end{document}